\documentstyle[12pt]{article}
\begin{document}

\vskip 1cm
\begin{center}
{\large{\bf A Note About Localized Photons on the Brane}}\\
\vskip 1cm
{\bf A. Khoudeir}\\
{\ Instituto de F\'{\i}sica, Universidad Nacional Aut\'onoma de M\'exico\\
Apdo. Postal 20-364, 01000 M\'exico D. F. M\'exico}\\

and

{\it Centro de Astrof\'{\i}sica Te\'orica, Departamento de F\'{\i}sica,
Facultad de Ciencias, Universidad de los Andes,\\
M\'erida, 5101,Venezuela.}\\

\end{center}

\begin{center}
email: adelkm@fenix.ifisicacu.unam.mx. adel@ciens.ula.ve
\end{center}

\begin{abstract}
A first order formulation for the Maxwell field in five dimensions is 
dimentionally reduced using the Randall-Sundrum mechanism. We will see 
that massive photons can not be localized on the brane.

\end{abstract}

In the Randall-Sundrum mechanism\cite{rs}, gravitons and scalar fields 
in $D = d + 1$ dimensions described by the Einstein-Hilbert 
and Klein-Gordon actions respectively, lead to gravitons and scalar 
fields localized on a d-dimensional brane. But, the usual (second order) Maxwell action can 
not yield photons trapped on the brane\cite{kss}. A solution of this dilemma 
was achieved in ref.\cite{lp} where the photons localized on the brane in 
four dimensions come from two antisymmetric tensors in five dimensions, 
satisfying a first order self-dual action\cite{tpvn}. 
Odd dimensional self-dual actions can be obtained from 
Kaluza-Klein dimentional reduction \cite{nvv}. 
Specifically, Lu and Pope\cite{lp} found that the 
bosonic sector of ungauged $N = 2$, $D = 4$ supergravity can be obtained 
from $N = 4$, $D = 5$ gauged supergravity. The supersymmetric extension 
was considered by Duff, et al.\cite{dls} . 
The generalization for higher antisymmetric fields (p-forms) gives  
some negative results. Indeed, if we consider the Randall-Sundrum ansatz 
for the metric
\begin{equation}
ds^2 = e^{-2k|z|}g_{mn(x)}dx^m dx^n + dz^2 ,
\end{equation}
where $m,n = 0,1,...,d$, together with the natural anzatz for a p-form
\begin{equation}
A_{M_{1}...M_{p}(x,z)} = A_{M_{1}...M_{p}(x)} ,
\end{equation}
then, the p-form is localized on the brane if 
\begin{equation}
p < \frac{d - 2}{2} .
\end{equation}
For $d = 4$ (five dimensions), we have $p < 1$ and only scalar fields 
($0$-form) can be trapped on the brane. But, a scalar field is dual to 
a $3$-form in five dimensions, which is not localized on the brane if the 
anzatz (2) is considered. In consequence, 
apparently, the Randall-Sundrum mechanism 
can not explain the dual equivalence of $p$-forms. However, 
Duff and Liu \cite{dl} have solved this dilemma. They found that the right anzatz 
for $p$-forms is given by 
\begin{equation}
A_{m_{1}...m_{p-1}z(x,z)} = 
e^{-2(p - \frac{d}{2})k|z|} A_{m_{1}...m_{p-1}(x)},
\quad A_{m_{1}...m_{p}(x,z)} = 0 .
\end{equation}
With this anzatz, the duality between $p$-form and $(D - p -2)$-form in the 
bulk implies the duality between $(p - 1)$-form and $(D -(p-1) -2)$-form 
on the brane. The criterion for consistency in this case is 
\begin{equation}
p > \frac{d}{2} . 
\end{equation}
If $d = 4$, we see that photons are excluded too. Thus, in five dimensions, 
p-forms with $p = 0$ and $p > 2$ can be trapped on the brane, ruled out 
the possibility of bounding photons (1-form) and its dual partner (2-form) 
on the brane\cite{dl}. 
The consistency with the Einstein equation will impose additional 
restrictions \cite{dl}. These results are 
based on the consideration of a second order action for p-forms.
If we try to modify the 
ansatz (2) for vectors, including an explicit dependence with $z$ 
in the following way: $A_{M} = (e^{-ak|z|}A_{m}, 0)$\cite{dls}, then the  action for massive vectors is obtained. 
In this note, we will see that not only massless vector fields 
but massive vector fields can not be localized on the brane 
in the framework of Randall-Sundrum dimentional reduction.

Let us start with the following first order action in five dimensions
\begin{equation}
S= -\frac{1}{12}\int d^4 x dz [\epsilon^{MNPQR}H_{MNP}F_{QR} + \sqrt{-g} g^{MQ}
g^{NR}g^{PS}H_{MNP}H_{QRS}],
\end{equation}
where $F_{MN} = \partial_{M}A_{N} - \partial_{N}A_{M}$. $H_{MNP}$ and $A_{M}$ 
are independent fields. We can introduce the dual of $H_{MNP}$ 
($= \frac{1}{2}\sqrt{-g}\epsilon_{MNPQR}f^{QR}$) and (6) becomes the usual 
first order formulation for the Maxwell action.
Their equations of motion are 
\begin{equation}
H^{MNP} = -\frac{1}{2\sqrt{-g}}\epsilon^{MNPQR}F_{QR}
\end{equation}
and 
\begin{equation}
\epsilon^{MNPQR}\partial_{N}H_{PQR} = 0 . 
\end{equation}

Substituting eq. (7) into action, we obtain the second order Maxwell action. 
On the other hand, eq. (8), can be solved (locally)
\begin{equation}
H_{MNP} = \partial_{M}B_{NP} + \partial_{N}B_{PM} + \partial_{P}B_{MN}
\end{equation}
and substituting in the action, it becomes the action for an antisymmetric field 
($B_{MN}$). In other words, the action (6) show us the dual equivalence between 
$A_{M}$ and $B_{MN}$ in five dimensions. 

Now, we apply the Randall-Sundrum anzatz for the metric (eq. (1)). The 
ansatz for the $H_{MNP}$ field is
\begin{equation}
H_{mnz(x,z)} = e^{-k|z|}b_{mn(x)}, \quad H_{mnp(x,z)} = 0 
\end{equation}
and for the vector field
\begin{equation}
A_{m(x,z)} = e^{-k|z|}A_{m(x)}, \quad A_{z(x,z)} = 0 .
\end{equation}

The following reduced action is obtained
\begin{equation}
S = \int dz e^{-2k|z|} \int d^4 x [-\frac{1}{4}\epsilon^{mnpq}b_{mn}F_{pq} - 
\frac{1}{4}\sqrt{-g_{4}}b_{mn}b^{mn} ].
\end{equation}
The (auxiliary) field $b_{mn}$ can be eliminated using its equation of motion 
\begin{equation}
b^{mn} = -\frac{1}{2\sqrt{-g_{4}}}\epsilon^{mnpq}F_{pq}
\end{equation}
and the Maxwell action trapped on the brane is obtained
\begin{equation}
S = \int dz e^{-2k|z|} \int d^4 x [ -\frac{1}{4}\sqrt{-g_{4}}F_{mn}F^{mn} ] .
\end{equation}
Although the Maxwell action appear on the brane, the ansatz is not consistent 
with the Einstein equation with cosmological constant 
($\Lambda = -6k^{2}$). Indeed, we have the Einstein equation 
\begin{equation}
R_{MN} - \frac{1}{2}g_{MN}R = g_{MN}\Lambda + T_{MN}
\end{equation}
where
\begin{equation}
T_{MN} = -\frac{1}{2}g^{PR}g^{QS}H_{MPQ}H_{NRS} + \frac{1}{12}
g_{MN}H_{PQR}H^{PQR} .
\end{equation}
Substituting the ansatz (1) and (10), we find for the $mn$ components
\begin{equation}
R_{4mn} - \frac{1}{2}g_{mn}R_{4} = g^{pq}F_{mp(A)}F_{nq(A)} - \frac{1}{4}
g_{mn}g^{pr}g^{qs}F_{pq(A)}F_{rs(A)} 
\end{equation}
where we have used eq. (13). Note that both sides of eq. (17) depend only 
of the coordinates $x$. Taking trace, we have $R_{4} = 0$. 
Then, the $zz$ component leads to an inconsistency
\begin{equation}
F_{mn}F^{mn} = 0. 
\end{equation}

Then, there is no consistency with the Einstein equation. 
This situation is similar to what happen in 
the usual Kaluza-Klein dimentional reduction if the zz component of 
the metric is $g_{zz} = 1$, instead of having $g_{zz} = \phi$. But 
the Randall-Sundrum ansatz does not admit any scalar field. 
Moreover, Duff and Liu \cite{dl} have shown that 
massless scalar and third rank antisymmetric fields, in five 
dimensions are the p-forms 
compatible with the Randall-Sundrum dimentional reduction when 
the coupling with gravity is considered.  
Furthermore, the ansatz $H_{mnp(x,z)} = 0$ is inconsistent with the equations of motion (7) and (8). Then, we must modify the anzatz for $H_{mnp(x,z)}$. 
This will lead to the action of a massive vector field on the brane, but 
the same inconsistency in the Einstein equations will be present. 
For instance, we choose 

\begin{equation}
H_{mnp(x,z)} = e^{-2k|z|}(\partial_{m}B_{np(x)} + \partial_{n}B_{pm(x)} + 
\partial_{p}B_{mn(x)}) \equiv e^{-2k|z|} h_{mnp(x)}, 
\end{equation}

where $B_{mn(x)}$ is a second rank antisymmetric field. 
The following reduced action on the brane is obtained

\begin{equation}
S = \int d^4 x dz e^{-2k|z|}[-\frac{1}{4}\sqrt{-g}F_{mn}F^{mn} 
-\frac{1}{12}h_{mnp}h^{mnp} - \frac{1}{4}\mu_{z}\epsilon^{mnpq}B_{mn}F_{pq}]
\end{equation}
where $\mu_{z} = ke^{-k|z|}$ is a mass parameter. We identify the 
Cremmer-Sherk action \cite{cs} on the brane, which describes massive vector fields in a gauge invariant way. This action can be obtained from 
Kaluza-Klein dimentional reduction \cite{adel} 

The mn components of the Einstein equation are now 
\begin{eqnarray}
R_{4mn} - \frac{1}{2}g_{mn}R_{4} &=& g^{pq}F_{mp(A)}F_{nq(A)} - \frac{1}{4}
g_{mn}g^{pr}g^{qs}F_{pq(A)}F_{rs(A)} \\ \nonumber
&-& \frac{1}{2}g^{pr}g^{qs}h_{mpq}h_{nrs} + \frac{1}{12}g_{mn}h_{pqr}h^{pqr}.
\end{eqnarray}

Taking trace, the following value of the scalar curvature is obtained 
\begin{equation}
R_{4} = \frac{1}{6}h_{mnp}h^{mnp}
\end{equation}
 
Using this value of $R_{4}$, the zz component of the Einstein equation 
yields the same inconsistency found previously, i.e., 
$F_{mn}F^{mn} = 0$. The same result is obtained if another ansatz for 
$H_{mnp}$ is considered, e.g.,
$H_{mnp} = f_{(z)}\sqrt {-g} \epsilon_{mnpq} A^{q}$.

Summarizing, we have considered a first order formulation for the 
Maxwell action in five dimensions and 
we have seen that massive photons can not be localized on the 
brane in a consistent way, using the Randall-Sundrum mechanism.

\begin{center}
{ACKNOWLEDGEMENTS}
\end{center}

The author would like to thank to Marti Ruiz Altaba for his hospitality 
at Instituto de F\'{\i}sica de la Universidad Nacional Aut\'onoma de 
M\'exico. Also, the author thanks Conicit-Venezuela for financial support.


\newpage


\begin{thebibliography}{60}

\bibitem{rs} L. Randall and R. Sundrum, Phys. Rev. Lett. {\bf 83} (1999) 4690.

\bibitem{kss} N. Kaloper, E. Silverstein and L. Susskind, 
{\bf hep-th/0006192}; A. Pomarol, Phys. Lett. {\bf B486} (2000) 153 and 
Phys. Rev. Lett. {\bf 85} (2000) 4004.

\bibitem{lp} H. Lu and C. N. Pope {\bf hep-th/0008050}. 

\bibitem{tpvn} P. K. Townsend, K. Pilch and P. van Nieuwenhuizen, Phys. Lett. 
{\bf B136} (1984) 38;  S. Deser and R. Jackiw, Phys. Lett. {\bf B139} 
(1984) 371.

\bibitem{nvv} H. Nastase, D. Vaman and P. van Nieuwenhuizen. Nucl. Phys. 
{\bf B581} (2000) 179.

\bibitem{dls} M. J. Duff, J. T. Liu and W. A. Sabra {\bf hep-th/0009212}.

\bibitem{dl} M. J. Duff and J. T. Liu {\bf hep-th/0010171}. 

\bibitem{cs} E. Cremmer and J. Scherk Nucl. Phys. {\bf B72} (1974) 117.

\bibitem{adel} A. Khoudeir Phys.Rev. {\bf D59} (1999) 027702.


\end{thebibliography}
\end{document}